\documentclass[floatfix,aps,twocolumn,superscriptaddress,prb]{revtex4-2}
\usepackage{graphicx}% Include figure files
\usepackage{dcolumn}% Align table columns on decimal point
\usepackage{bm}% bold math
\usepackage[T1]{fontenc}
\usepackage[latin9]{inputenc}
\usepackage{textcomp}
\usepackage{amsmath}
\usepackage{mathrsfs}
\usepackage{graphicx}
\usepackage{amssymb}
\usepackage{units}
\usepackage{xcolor}
\usepackage{float}

\newcommand{\ivan}[1]{\textcolor{blue}{#1}}
\def\sezioni{1}

\begin{document}

\title{Polaron spectroscopy of a bilayer excitonic insulator}

\author{Ivan Amelio}
\affiliation{
Institute of Quantum Electronics ETH Zurich, CH-8093 Zurich, Switzerland}
\author{N.\ D.\ Drummond}
\affiliation{Department of Physics, Lancaster University, Lancaster LA1 4YB, United Kingdom}
\author{Eugene Demler}
\affiliation{Institute for Theoretical Physics, ETH Zurich, 8093 Zurich, Switzerland}
\author{Richard Schmidt}
\affiliation{Max-Planck-Institut f\"{u}r Quantenoptik, 85748 Garching, Germany}
%\affiliation{Munich Center for Quantum Science and Technology, 80799 M\"{u}nchen, Germany}
% \affiliation{Center for Complex Quantum Systems, Department of Physics and Astronomy, Aarhus University, 8000 Aarhus C, Denmark}
\affiliation{Institut f\"ur Theoretische Physik, Universit\"at Heidelberg, Philosophenweg 16, 69120 Heidelberg, Germany}
\author{Atac Imamoglu}
\affiliation{
Institute of Quantum Electronics ETH Zurich, CH-8093 Zurich, Switzerland}

\begin{abstract}
Recent advances in fabrication of two dimensional materials and their moir\'e heterostructures have opened up new avenues for realization of ground-state excitonic insulators, where the structure spontaneously develops a finite interlayer electronic polarization.
We propose and analyze a scheme where an optically generated intralayer exciton is screened by excitations out of the excitonic insulator to form interlayer polarons. Using Quantum Monte-Carlo calculations we first determine the binding energy of the biexciton state  composed of inter- and intralayer excitons, which plays a central role in understanding polaron formation. We describe the excitations out of the ground-state condensate using  BCS theory and use a single interacting-quasiparticle-pair excitation Ansatz to describe dynamical screening of optical excitations. Our predictions carry the hallmarks of the excitonic insulator excitation spectrum and show how changing the interlayer exciton binding energy by increasing the layer separation modifies the optical spectra. 
\end{abstract}

\date{\today}
\maketitle

\if\sezioni1
\section{Introduction}
\fi

An excitonic insulator (EXI) is a phase of matter where the ground-state features
bound electron-hole pairs~\cite{keldysh1964,keldysh1968collective}.
This is most easily realized in bilayer structures where the lowest energy conduction band (CB) state of one layer is tuned near resonance with the highest energy valence band (VB) state of the other layer. Introduction of insulating layers in between the layers suppresses tunnel coupling, thereby ensuring
separate charge conservation in the two layers~\cite{zhu1995exciton}. Exciton formation corresponds to the 
binding of electron and hole pairs due to Coulomb attraction.
{Due to the aligned dipole moments, such ground-state excitons are a promising candidate to mediate interactions between itinerant electrons (holes) in the CB (VB),} providing a  platform for the physics of Bose-Fermi mixtures \cite{enss2009superfluidity,pierce2019few,wang2022light-induced}, potentially supporting superconductivity~\cite{laussy2010expol,kinnunen2018induced,cotlet2016superconductivity}.
Recently, evidence for the formation of ground-state excitons in bilayer transition metal dichalcogenides (TMDs) in the absence of a magnetic field has been reported using capacitance measurements~\cite{ma2021strongly}.

In this Letter, we 
propose optical spectroscopy as a probe of excitonic insulators. We particularly focus on an electric field tunable MoS$_2$/hBN/WSe$_2$ heterostructure where the conduction band (CB) of MoS$_2$ can be tuned into resonance with the valence band (VB) of WSe$_2$~\cite{wu2015theory}. We assume that an intralayer exciton (X) is injected by resonant light absorption, which in turn acts as a quantum impurity that can bind to interlayer excitons (IXs) in the ground-state.
Polaron spectroscopy has already proved to be an invaluable tool to characterize many-body states in TMD mono- and bilayers~\cite{sidler2017fermi, ravets2018polaron,shimazaki2020strongly, smolenski2021signatures}.

{In the limit where the EXI is described by a dilute Bose gas of IXs, the physics of the mobile impurity physics may be regarded as a Bose polaron problem.} 
Since the binding energy between the impurity and the bosons plays a central role in understanding the polaron spectrum, we compute the binding energy of this inter-intra layer biexciton X-IX by a 4-body Quantum Monte-Carlo (QMC) calculation. For large interlayer distances, the biexciton wavefunction and energy  approach that of an intralayer trion (T) loosely bound to a hole in the other layer.

However, the Bose polaron description does not generally apply to our system. In the first place, for increasing interlayer distances the IX binding energy becomes comparable to the trion binding energy and the internal structure of the IX plays a role.
Moreover, at larger chemical potentials, 
a description of the ground-state in terms of tightly bound, point-like bosons  is inadequate, but rather  pairing involves fermions close to the Fermi surface.

To fully take into account the microscopic fermionic nature of the system, we model the EXI using the mean-field BCS formalism~\cite{jerome1967excitonic,keldysh1968collective,comte1982exciton}, while treating the intralayer exciton as a rigid mobile impurity.
The polaron spectra are computed using a generalization of the Chevy Ansatz, where two interacting fermionic quasiparticles are excited and scatter off the mobile impurity.
This analysis recovers the expectation that the energy of the attractive polaron at low IX densities is determined by the X-IX binding energy. Moreover, we find that the gap in the quasiparticle spectrum hampers the  transfer of the oscillator strength from the repulsive to the attractive branch. Interestingly, when the IX binding energy is comparable to the quasiparticle gap,
we  predict the emergence of a third peak, associated with an excited state of the X-IX complex.
Polaron spectroscopy of an EXI  carries clear signatures of interlayer pairing and may provide a direct estimate of the quasiparticle pair excitation gap. Moreover, potential valley polarization of the EXI~\cite{wu2015theory} could be easily assessed in polarization resolved spectroscopy.

On a technical level, the generalized Ansatz we use has been previously implemented in the context of 3D atomic Fermi superfluids~\cite{yi2015polarons,hu2022crossover,amelio2022two-dimensional}
and has allowed for interpolation
between different regimes, including the BEC-BCS crossover for varying densities.
To get a better understanding of the variational subspace, 
we analyze the elementary neutral excitations of the EXI; we show that even though the gapless Goldstone branch is not captured at the single-excitation level, all other collective modes, including the ones referred to in the literature as Higgs~\cite{xue2020higgs} or Bardasis-Schrieffer~\cite{sun2020bash} modes, are well reproduced. In computing the elementary excitations, we extend the geometric approach of \cite{guaita2019gaussian, hackl2020} to  BCS theory, where the BCS state is treated as a Gaussian state and the collective modes as fluctuations on the Gaussian manifold.
Moreover, we do not assume contact interactions, but rather work with realistic bilayer Keldysh~\cite{rytova1965,keldysh1979coulomb,danovich2018} and exciton-electron~\cite{fey2020theory} interactions, 
marking a key difference with respect to atomic superfluids where interactions are contact-like and   the polaron behavior was studied as a function of the 3D scattering length tuned through a Feschbach resonance.

\if\sezioni1
\section{Few-body binding energies}
\fi

\begin{figure*}[t]
    \centering
    \includegraphics[width=0.33\textwidth]{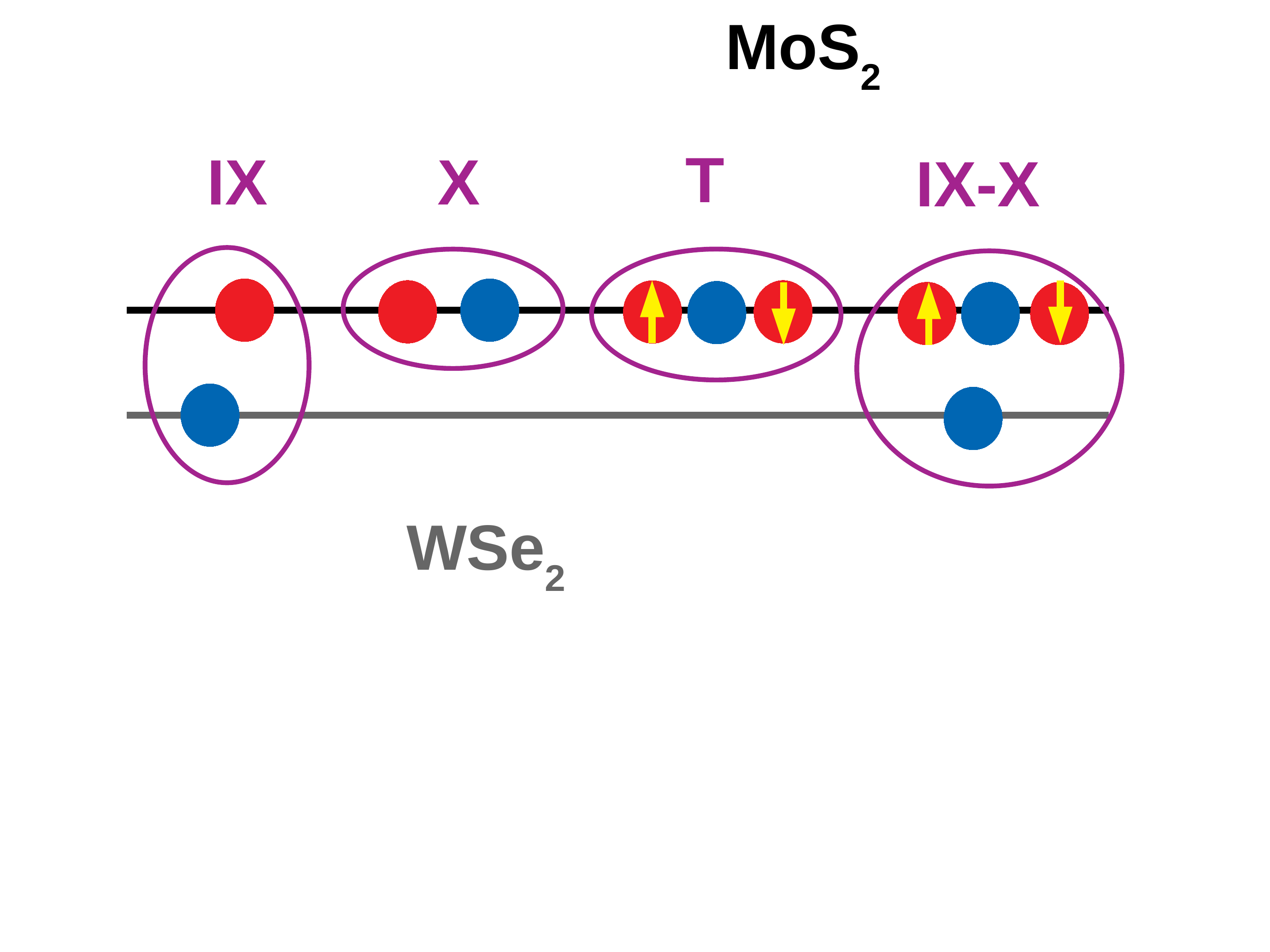}
    \includegraphics[width=0.66\textwidth]{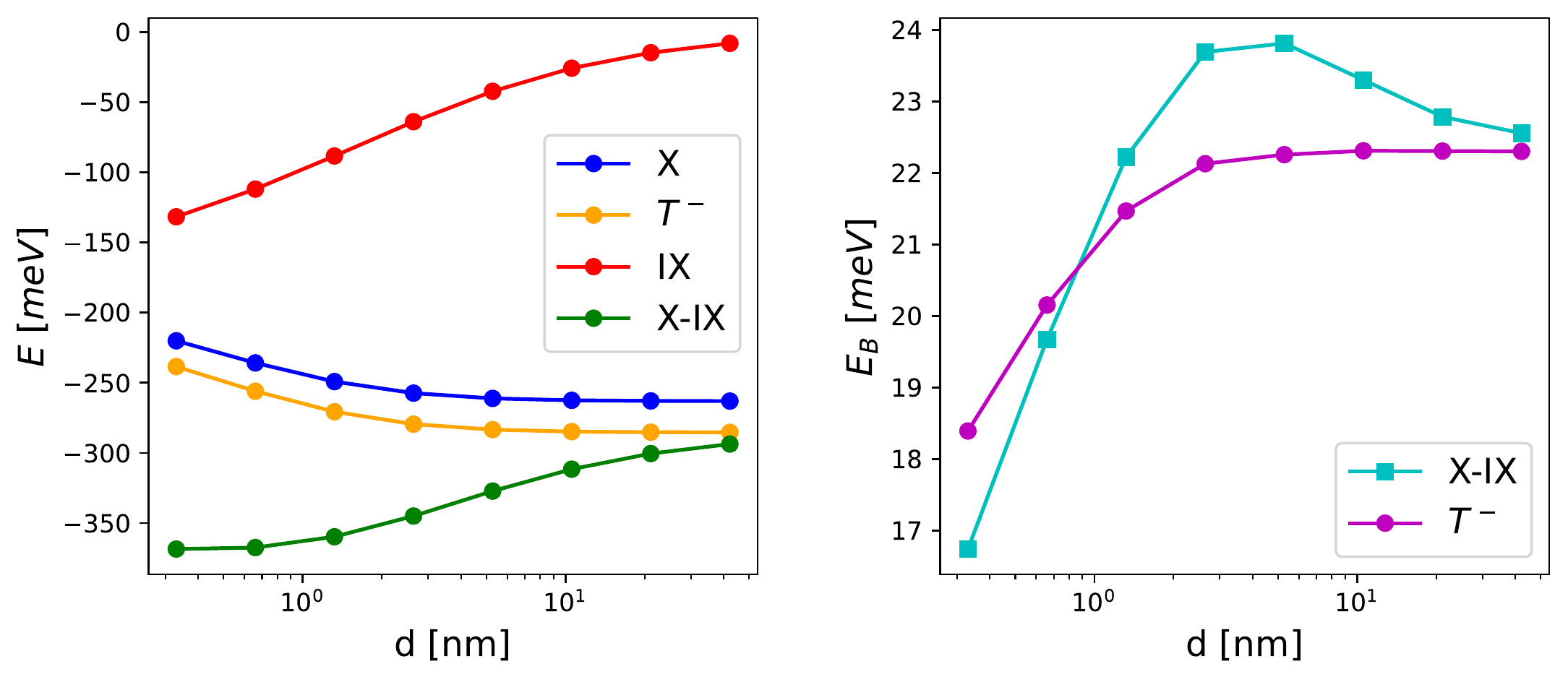}
    \caption{(a) Sketch of the X, IX, T, X-IX complexes in terms of electrons and holes (red and blue circles respectively).
    Energies (b) and binding energies (c) for different complexes and as a function of the interlayer distance $d$. These results are the output of QMC calculations with parameters suitable for a MoS$_2$/hBN/WSe$_2$ heterostructure.
 }
    \label{fig:Eb}
\end{figure*}

Experiments in charge-tunable TMD monolayers have established that the dominant resonances in the optical excitation spectra can be identified as attractive and repulsive polarons (AP and RP)\@. In the limit of vanishing doping, the AP resonance  energy approaches that of a T~\cite{sidler2017fermi}. Similarly, since the ground-state of the bilayer system in the small density BEC regime consists effectively of tightly bound IXs, we expect that the X-IX energy determines the position of the AP resonance. Determining the binding energies of X, IX, T and X-IX, is therefore key to understand the optical excitation spectra as a function of interlayer separation.

We compute the binding energies of different complexes in a MoS$_2$/hBN/WSe$_2$ heterostructure using the diffusion QMC method as implemented in the \textsc{casino} package~\cite{needs2020variational}.
prisedcomIn particular, as sketched in Fig.~\ref{fig:Eb}a,  we calculate the binding energies of the MoS$_2$ X, the IX made of an electron in  MoS$_2$ and a hole in  WSe$_2$, the T in  MoS$_2$ and the X-IX\@.
Our calculation is an extension of Refs.\ \onlinecite{szyniszewski2017binding} and \onlinecite{mostaani2017diffusion} to a bilayer system; importantly, the trial wavefunctions decay exponentially at large separations and satisfy the cusp conditions for the Keldysh potential~\cite{mostaani2017diffusion}. The Jastrow function  includes smoothly truncated polynomial expansions in interparticle distances~\cite{drummond2004jastrow}.

The QMC approach can only treat the intraband sectors of the Coulomb operator. Even though the electron-hole exchange terms do modify the binding energy of T, and X-IX~\cite{courtade2017charged,mostaani2017diffusion}, their contribution should be small as compared to the actual binding energy. Based on that we expect QMC calculation to provide a good estimate of the binding energies.

We take $m_\text{e} =m_\text{h}=0.55 m_0$ for the CB and VB  masses in MoS$_2$, and  $m_\text{h}=0.40 m_0$ for the VB in WSe$_2$. These values match the reduced masses reported in the quantitative investigation by Goryca \textit{et al.}\ \cite{goryca2019revealing}. {We neglect the less known mass ratio imbalance, having verified that the binding energies are not very sensitive to it~\cite{courtade2017charged}.}
As detailed in SI.A, we use the bilayer Keldysh potential~\cite{danovich2018} for electron-hole and electron-electron interactions:  
We take the relevant screening lengths from Ref.~\cite{goryca2019revealing}, and use $\epsilon=4.5$ for the dielectric constant of the hBN environment. With these parameters, for a MoS$_2$ monolayer, we obtain 217 meV for the X binding energy and 18 meV for the trion, which are in good agreement with the experimental values in encapsulated samples~\cite{roch2019spin-polarized}.

{Figure~1.b shows the calculated energy of the few-body complexes as a function of the interlayer distance $d$, having set as reference to zero the energies of the band edges. We express $d$ in units of the thickness of a single hBN layer ($L_1 = 0.33$ nm), and plot X, T, IX,X-IX for $d = 1,2,4,\dots,128 \times L_1$. We note that the MoS$_2$ exciton and trion energies also depend on $d$, since the WSe$_2$ monolayer contributes to screening. 
The binding energies of T and X-IX, extracted from the energies depicted in Fig.~1b are plotted in Fig.~\ref{fig:Eb}.c; here the binding energy $E_b$ for X-IX is given by
$E_b = E_{\rm X} + E_{\rm IX} - E_{\rm X-IX}$.
For large interlayer distances yielding $E_X \gg E_{IX}$, the X-IX state can be described as a strongly bound negatively charged intralayer trion T loosely bound to a hole in the WSe$_2$ layer (see SI.B for the probability distribution functions).}

\if\sezioni1
\section{Many-Body Model}
\fi

To analyze optical excitation spectrum in the presence of an EXI, we assume that the X can be treated as a point-like  quantum impurity with no internal degrees of freedom. We also assume that IXs are spin-valley polarized and a few-layer-thick hBN separates MoS$_2$ and WSe$_2$ layers, ensuring that there is no electronic moir\'e potential. For simplicity, we consider the particle-hole symmetric case, with equal electron and hole masses ($m_\text{e} = m_\text{h} = m$) and set chemical potentials $\mu_\text{e} = \mu_\text{h} = \mu$.
In practice, $\mu_\text{e}, \mu_\text{h}$
can be tuned either by connecting the two layers to different reservoirs to form a biased junction or by applying a normal electric field $E_z$.
In a recent experiment \cite{ma2021strongly} both mechanisms have been employed to overcome the semiconductor band gap. Our theory can be equally applied to both scenarios.

The total system Hamiltonian is given by
\begin{equation}
    H = H_\text{el} + H_\text{imp} + H_\text{W}.
\end{equation}
Denoting CB and VB electrons respectively as $a \equiv a_c, b \equiv a_v$, the electronic Hamiltonian reads
\begin{multline}
    H_\text{el} = \sum_k \varepsilon_k (a^\dagger_k a_k +  b_k b^\dagger_k)
    + \frac{1}{2A}
    \sum_{kpq} U(q) (a^\dagger_{k+q} a^\dagger_{p-q}
    a_p a_k
    +
    \\
    +b_{k+q} b_{p-q}
    b^\dagger_p b^\dagger_k )
    - \frac{1}{A}
        \sum_{kpq} V(q) a^\dagger_{k+q} b_{p+q}
    b^\dagger_p a_k 
\end{multline}
where $\varepsilon_k = \frac{k^2}{2m} - \mu$. The second term represents intralayer Coulomb repulsion while the last term is the interlayer  interaction.
More precisely, we take $U,V$ as the bilayer Keldysh interaction potentials:
\begin{equation}
    U(q)=
    V(q)
    \left[
    (1+r_* q)e^{qd} - 
    r_* q e^{-qd}
    \right],
    \label{eq:Keldysh_intra}
\end{equation}
\begin{equation}
    V(q)=
    \frac{2\pi}{q}
    \left[
    (1+r_* q)^2 e^{qd} - 
    r_*^2 q^2 e^{-qd}
    \right]^{-1}.
    \label{eq:Keldysh_inter}
\end{equation}
Here we set 
$\frac{e^2}{4\pi \epsilon_0 \epsilon}=1$, assume the same screening length $r_*$ for the two TMDs and use $V(q) = \int dx \ e^{-iqx} V(x)$ for the Fourier transform.

The impurity Hamiltonian is
$H_\text{imp} = \sum_q \epsilon^I_q x^\dagger_q x_q$
with 
$ \epsilon^I_q
= E_X + \frac{q^2}{2M}
-i \gamma (q) $. {Here, $M = 2m$ and $\gamma (q)$ denotes the momentum-dependent X radiative decay rate.}

The general form of $H_\text{W}$ describing the coupling between the  impurity (X) in MoS$_2$ and the CB and VB electrons that make up the IX is given by
\begin{equation}
    H_\text{W} = \frac{1}{A} \sum_{kpq} W_a(q) x^\dagger_{p-q} x_p a^\dagger_{k+q} a_k
    -
    W_b(q) x^\dagger_{p-q} x_p    b_k b^\dagger_{k+q}.
\end{equation}
To model  the exciton-electron scattering, we adopt the effective  potential
\begin{equation}
    \tilde{W}(r) = \frac{W_0}{(r^2 + a_X^2)^2},
\end{equation}
which correctly describes the polarization of the exciton by the electron for $r \gg a_X$~\cite{fey2020theory}. We note that the Fourier transform of $ \tilde{W}(r)$ gives
$ W(q) = 
    \int d^2r \  \tilde{W}(r) e^{-iqr}
    = W_0 \frac{\pi}{a_X} q K_1(a_Xq), $ 
with $K_1$ denoting the modified Bessel function, ensuring 
$W(q \to 0) \to \frac{\pi}{a_X^2}W_0$.

The constant $W_0$ is chosen to ensure that $\tilde{W}(r)$ supports a single bound trion state with binding energy $\sim 20$ meV. Due to the short-ranged nature of the exciton-electron and exciton-hole interactions, we set $W_b=0$ since it describes the interaction between spatially separated excitons and holes that have vanishing wavefunction overlap. Including $W_b \neq 0$ is straightforward and we have verified that it does not lead to qualitative changes.

\if\sezioni1
\section{BCS-Chevy approach}
\fi

\begin{figure*}[t]
    \centering
    \includegraphics[width=0.33\textwidth]{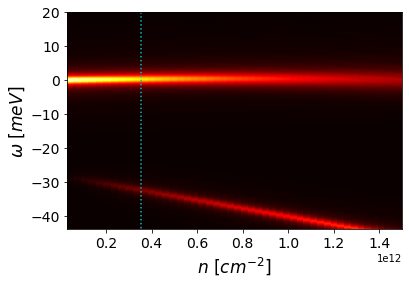}
    \includegraphics[width=0.33\textwidth]{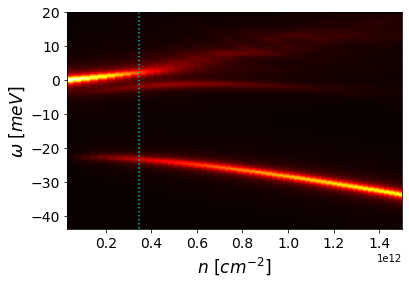}
        \includegraphics[width=0.30\textwidth]{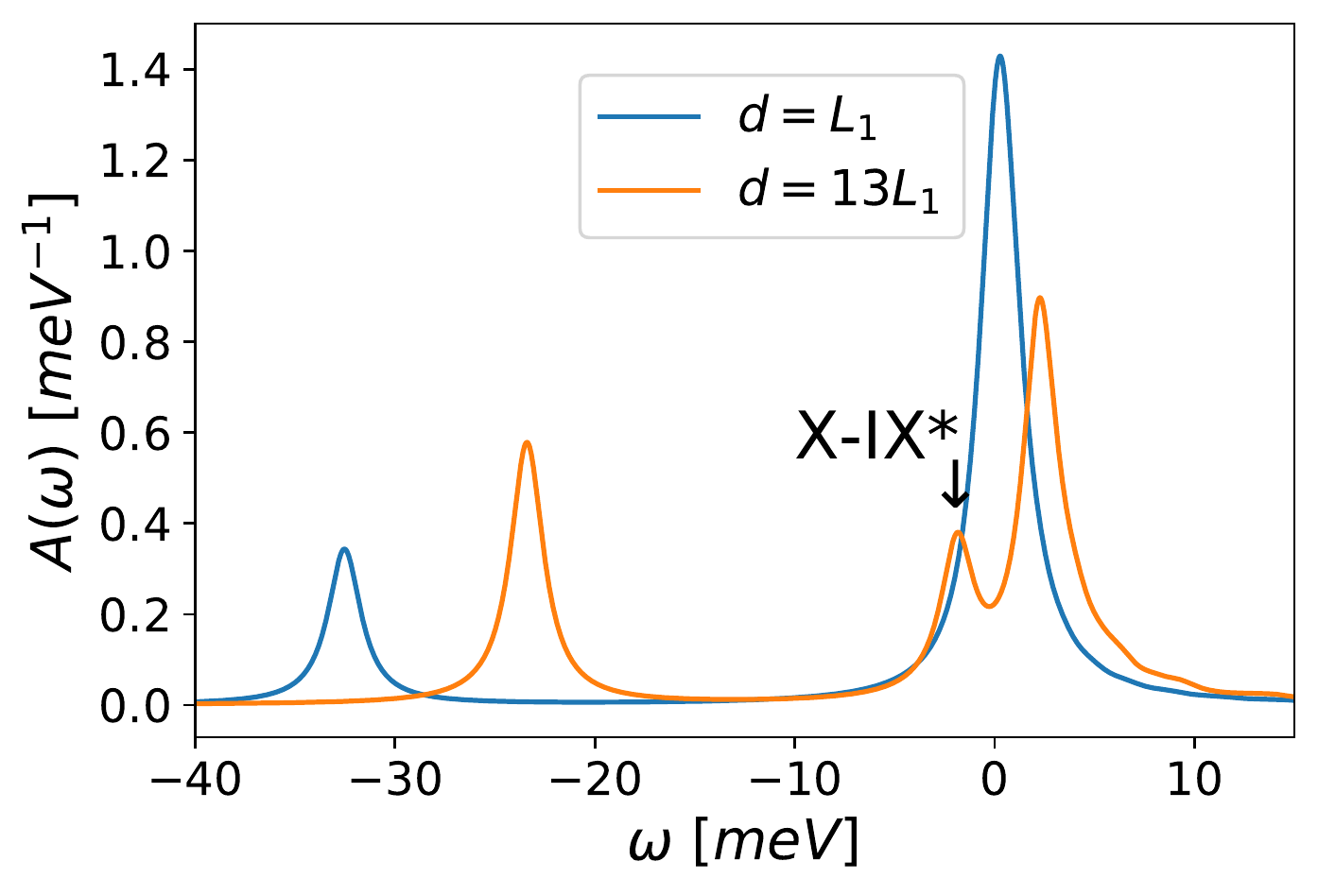}
    \caption{Polaron spectra as a function of the density for $d= L_1$ (a) and $d= 13 L_1$ (b).
    (c) Slice at $n \simeq 3.5 \cdot 10^{11}$cm$^{-2}$ (cyan dotted line in (a,b)), to highlight the high visibility of the RP for strong pairing (small $d$) as well as of the X-IX$^*$ peak at large
    $d$.}
    \label{fig:spectra_nw}
\end{figure*}

Even in the absence of an impurity, finding the many-body ground state (GS) of $H_\text{el}$ represents a formidable task.
A rich zero temperature phase diagram,  
which includes Wigner crystal, electron-hole plasma, exciton and biexciton condensate phases,
has been obtained in QMC calculations~\cite{depalo2002excitonic,maezono2013,depalo2022quadriexcitons}.
In this work, we are  interested in the excitonic phase and   resort to a BCS variational approach,
which can interpolate between the Bose-Einstein condensate (BEC) regime, {with $-2\mu$ close to the  IX binding energy, and the BCS regime at larger $\mu$.
This approach has  been applied previously to the excitonic insulator with  Coulomb   potential~\cite{jerome1967excitonic,keldysh1968collective,comte1982exciton,zhu1995exciton,wu2015theory}.}

To proceed, we introduce fermionic quasiparticle operators $\alpha_k$ and $\beta_k$ defined by: 
\begin{equation}
    a_k = u_k \alpha_k + v_k \beta_k 
    \ , \ \ \ \ \ \
    b_k = u_k \beta_k - v_k \alpha_k  \ .
\end{equation}
We express the GS as the vacuum of $\beta$ and the completely filled state of $\alpha$ quasiparticles (see also Fig. S2  of SI.C):
\begin{equation}
    \alpha_k^\dagger \alpha_k |\text{EXI}\rangle = |\text{EXI}\rangle
    \ , \ \ \ \ \ \
    \beta_k |\text{EXI}\rangle = 0 .
\end{equation}
We have  
$ n_k \equiv \langle a^\dagger_k a_k\rangle = \langle b_k b^\dagger_k \rangle = u_k^2$, so that the carrier density in each layer is $n= \frac{1}{A}\sum_k n_k$; in the BEC regime, $n$ is also the density of IXs. The  coefficients that define the new fermionic quasiparticles can be written in terms of the variational parameters $\theta_k$ as $u_k = \cos\theta_k, v_k = \sin\theta_k$, with the saddle point condition 
\begin{equation}
    \tan2\theta_k = -
    \frac{\Delta_k}{\xi_k},
\end{equation}
where 
$ %\begin{equation}
\Delta_k = \frac{1}{A}\sum_{k'} V(k-k') u_{k'} v_{k'},    
$ %\end{equation}
is the gap function and
$ %$\begin{equation}
    \xi_k = \varepsilon_k + 2\pi e^2 d n
    - \frac{1}{A}\sum_{k'} U(k-k') n_{k'}
$ %\end{equation}
is the Hartree-Fock dispersion.
Exciting a  quasiparticle costs an energy
$E_k = \sqrt{\xi_k^2 + \Delta_k^2}$, meaning that
$\langle \alpha^\dagger_k H_\text{el} \alpha_k \rangle - 
    \langle  H_\text{el} \rangle =
    \langle \beta_k H_\text{el} \beta^\dagger_k \rangle - 
    \langle  H_\text{el} \rangle = E_k$.

The analog of the Chevy Ansatz~\cite{chevy2006universal}
is constructed by dressing the impurity with excited quasiparticle pairs~\cite{yi2015polarons,hu2022crossover}: 
\begin{equation}
    | \Psi \rangle 
    = \left\{
    \psi_0 x^\dagger_0
    + \frac{1}{A}
    \sum_{kQ} \psi_k(Q) 
    \beta^\dagger_{k+Q} \alpha_k x^\dagger_{-Q}
    \right\}
    |\text{EXI} \rangle.
    \label{eq:Chevy_ansatz}
\end{equation}
The Schr\"odinger equation restricted to the variational subspace is then
\begin{widetext}
\begin{equation}
    i\partial_t \psi_0 = W_a(0) n \ \psi_0 + \frac{1}{A} \sum_{kQ} \langle  x_0 H_\text{W}
    \beta^\dagger_{k+Q} \alpha_k x^\dagger_{-Q}
    \rangle \ \psi_k(Q)
    \label{eq:ChevySchoedinger_0}
\end{equation}
\begin{multline}
    i\partial_t \psi_k(Q) = [E_k + E_{k+Q} + \epsilon^I_Q +  W_a(0) n] \ \psi_k(Q) +  A \langle  x_{-Q} \alpha_k^\dagger  \beta_{k+Q}  
    H_\text{W}
    x_0^\dagger \rangle \ \psi_0
    + 
    \\
    +
     \sum_{k'Q'} \langle  x_{-Q} \alpha_k^\dagger  \beta_{k+Q}  [H_\text{el} - H_\text{BCS} + H_\text{W} - W_a(0) n]
    \beta^\dagger_{k'+Q'} \alpha_{k'} x^\dagger_{-Q'}
    \rangle \ \psi_{k'}(Q').
        \label{eq:ChevySchoedinger_kQ}
\end{multline}
\end{widetext}

The explicit expressions of  the matrix elements are given in  SI.E.
We emphasize that in deriving Eqs.\ (\ref{eq:ChevySchoedinger_0}) and (\ref{eq:ChevySchoedinger_kQ}), we did {\em not} linearize the electronic Hamiltonian $H_\text{el}$ to the Bardeen-Cooper-Schrieffer (BCS) form
$H_\text{BCS} = \langle  H_\text{el} \rangle
    +\sum_k E_k  \alpha_k \alpha^\dagger_k
    +     \sum_k E_k \beta^\dagger_k \beta_k$.
As a consequence, interactions between the quasiparticles $\alpha_k$ and $\beta_k$ are captured~\cite{yi2015polarons}.
As shown in SI.H, it is evident that a formalism that neglects quasiparticle interactions, cannot describe the X-IX bound state.
Such an approximation can be adequate for  cold atom problems~\cite{hu2022crossover}, but would drastically fail in the system we are analyzing. We also remark that Eqs.~(\ref{eq:ChevySchoedinger_0}) and (\ref{eq:ChevySchoedinger_kQ}) reduce to a redundant description of the usual single component Fermi polaron~\cite{chevy2006universal,combescot2007normal,parish2011polaron-molecule, schmidt2012fermi} when the interlayer Keldysh interactions are switched off. In this case one has $u_k = \Theta(k_\text{F} - k), v_k = \Theta(k - k_\text{F})$.

In the Appendix we provide a more detailed discussion of the advantages and limitations of the Chevy approach. In particular,  the spectrum of the single particle-hole excitation sector of the fermionic system is compared with the Anderson-Bogoliubov spectrum. The latter is computed treating the BCS state as a Gaussian state and the collective modes as fluctuations on the Gaussian manifold.
It turns out that most of the neutral excitations modes of the systems are well captured already at the level of a single particle-hole excitation, including the Higgs and Bardasis-Schrieffer modes.
The Goldstone-like mode presents instead the main challenge since in this sector it is gapped and disperses parabolically. However, in the very low density limit, the gap and the inverse healing length are much smaller than the X-IX binding energy and  inverse X-IX radius, so that the correction to the polaron spectra would be small.

\if\sezioni1
\section{Results}
\fi

\begin{figure}[t]
    \centering
    \includegraphics[width=0.49\textwidth]{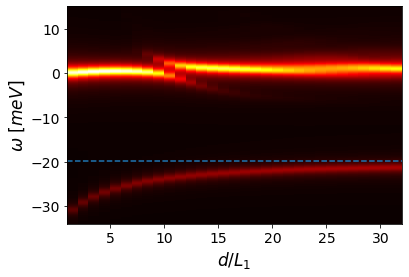}
    \caption{Polaron spectra  at the density $n \simeq 2\cdot 10^{11}$,
    as a function of the  interlayer separation.
    The X-IX$^*$ peak lowers in energy for larger $d$ and gains weight when crossing the repulsive polaron to disappear in the AP-RP gap for large $d$.
    At large $d$ we are in the Fermi polaron regime and the AP gets closer and closer to the energy of the monolayer trion (dashed cyan line).
    }
    \label{fig:track_XIX}
\end{figure}

We diagonalize the Chevy-Schr\"odinger equations (\ref{eq:ChevySchoedinger_0}-\ref{eq:ChevySchoedinger_kQ}) numerically; when computing the spectral function, the weight of each eigenmode  is given by the quasi-particle weight $|\psi_0|^2$.  The line broadening of the bare X is set to 
%a full width at half maximum of
$\gamma(0) = 1$ meV for zero exciton momentum  and zero otherwise.
In the calculations, we choose the electron and hole masses to be $m=0.5m_0$. Details on the numerical scheme are given in SI.F.

We compute the  polaron spectra as a function of $n_{IX}$ and $d$ as these are the experimentally tunable parameters.
For each parameter set we compute the mean-field GS and use the corresponding $u_k,v_k,E_k$'s in the Chevy-Schr\"odinger equations.  Since the Fermi polaron regime is recovered for $d\to \infty$, comparing spectra at small and large $d$  provides insight on the effect of pairing.

\if\sezioni2
\ivan{remove enumerate}
\begin{enumerate}
\fi
    \if\sezioni2
  \item 
  \fi
  Many important features can be observed in Fig.~\ref{fig:spectra_nw}, 
  which shows the calculated optical excitation spectra as a function of $n_{IX}$ for two extreme values of $d$, namely $d=L_1$ in (a) and $d=13L_1$ in (b). In both cases, we find that the principal spectral features consist of an AP branch that gains in strength and red shifts with increasing $n_{IX}$, while the accompanying RP branch blueshifts, broadens and loses oscillator strength. It is the small but discernible deviations that we discuss below that contains the signatures of EXI.

  First, at small IX density (BEC limit) we observe an AP branch originating from the X-IX bound state. As we argued above, this resonance is bright and acquires oscillator strength with increasing density.
    With the $W_a$ we used, the binding energy of the trion in the monolayer limit is around 20 meV, while for the X-IX at
    $d=L_1$ we have
    around 30 meV\@. 
    We remark  that these estimates are not directly comparable with the more reliable 
    4-body
    QMC computations of Fig.~\ref{fig:Eb}.c, since here the X is assumed to be point-like. In particular, the screening of the optically excited electron-hole pair forming the X by the IX is not accounted for, explaining the much larger binding energy of the X-IX\@.
    Since $E_T$ can be  directly measured by selectively doping only the electron layer, the presence of this peak at $E_{X-IX} \neq E_T$ would provide a direct evidence that the probe exciton is interacting with the paired IX\@. On the other hand, notice that for some interlayer distances $E_{X-IX} \simeq E_T$, so one cannot use the inverse argument to rule out the EXI.
    It is indeed likely that the small jump in the AP line measured in \cite{gu2022dipolar} is indeed due to the slightly different binding energy of the T and X-IX, though for a different kind of excitonic insulator.

 Another interesting feature,
highlighted in Fig.~\ref{fig:spectra_nw}.c, 
  emerges when we compare the RP branch for the two layer separations: the RP for $d=1L_1$ remains much brighter than its $d=13L_1$ counterpart for $n_{IX} \le 1 \times 10^{12}$~cm$^{-2}$.
 Two facts  play together here: first, in the limit of strong IX binding the electronic scattering states are  shifted to higher energies due to the  large pairing gap; second,  for a smaller binding energy the oscillator strength transfer occurs for smaller densities.
Similarly, the blueshift of the RP at 8 hBN spacers is around twice the blueshift with a hBN monolayer.
It may be interesting to make a comparison with  polarons on top of an incompressible state, as recently studied using DMFT in a homobilayer model~\cite{mazza2022arxiv}.

    Next we move on to another observation. A trion state may in principle exist at energy $\omega_T \sim 2E_0 - E_T$, where in the BEC limit $2E_0 = |E_{IX}|$ is the quasiparticle gap; this estimate corresponds to breaking up an IX to build a T plus a free hole. This state is not observed for small $d$. Instead, for larger interlayer distance such that $E_{IX} \lesssim  E_T$, a third peak appears in the spectrum, with sizable oscillator strength as it crosses the RP\@. The shift of this peak with $d$ is studied in Fig.~\ref{fig:track_XIX}. Inspection of the wavefunction (Fig. S5 of  SI.G) shows that, rather than a T plus a free hole, this is an excited $2s$ IX bound to the impurity or equivalently a T with a hole bound in a $2s$ wave.
   Notice that, with a realistic choice of parameters, the first excited state of the intralayer 3-body problem lies a few hundred meV higher in energy than the trion ground state~\cite{fey2020theory}, 
   so that quite generally a  third peak needs to be related to the excitation of the hole in the other layer.  
   Therefore, the position of this X-IX$^*$ peak allows for a direct estimate of the quasiparticle gap $2E_0 \sim E_{X-IX^*} - E_{X-IX}$.
   Notice that in the experiment \cite{ma2021strongly} the IX binding energy is estimated to be 25 meV, quite comparable with the typical trion binding energies: 
   this means that the secondary peak may be observed in existing platforms.
Moreover, it entails that polaron formation for the IXs in  \cite{ma2021strongly} requires a description that keeps into account the fermionic nature of their constituents, like the one provided here.  
    
\if\sezioni2
  \item 
  \fi

\if\sezioni2
\end{enumerate}
\fi

\if\sezioni1
\section{Conclusion and discussion}
\fi

In summary, we used a  generalized Chevy Ansatz [Eq.~(\ref{eq:Chevy_ansatz})]  to analyze the dressing of an intralayer exciton by the quasiparticle pair excitations  of a ground-state interlayer excitonic insulator.
The resulting polaron spectra carry clear signatures of interlayer pairing.

It was shown that a mean-field analysis which takes into account two valleys predicts  that the interlayer excitonic insulator would be valley polarized~\cite{wu2015theory,conti2020doping}. We have confirmed that this prediction remains true for the bilayer Keldysh potential (see SI.D). The presence of such spontaneous valley polarization could be detected using circularly polarized excitation of intralayer excitons: Since binding of inter- and intralayer excitons that we analyzed is primarily mediated through exciton-electron interactions within the same layer, we expect valley polarization of interlayer excitons with the bound electron in K'-valley, to lead to a valley-polarized attractive polaron resonance. However, we emphasize that such an experiment cannot rule out the possibility of an excitonic insulator where the electron and the hole reside in different valleys. 

A natural extension of our work would be the calculation of the polaron spectra at finite temperatures to demonstrate that optical polaron spectroscopy could be used to detect the transition from a thermal gas of excitons to a quasi-condensate. Recent experiments demonstrated that clear polaronic signatures of condensate formation indeed exist in three-dimensional ultracold BECs tuned across the critical temperature~\cite{yan2020bose}. Development of a formalism that goes beyond the zero-temperature mean-field description we used remains an open problem. 

Another exciting and timely extension of our formalism would be to analyze the optical signatures of interlayer exciton ground states in moir\'e heterostructures. In addition to s-wave paired ground-state moir\'e excitons that have been experimentally observed~\cite{zhang2022correlated,gu2022dipolar}, recent theoretical work proposed the possibility of spontaneous $p + i p$ exciton formation leading to quantum anomalous Hall effect~\cite{xie2022topological,dong2022excitonic,xie2022strongly}, or the emergence of fractional quantum Hall state of excitons~\cite{zhang2022su4}. Here, the use of polaron spectroscopy may help differentiate different competing ground states in these systems.

\if\sezioni1
\section{Acknowledgments}
\fi

We are grateful to Clemens Kuhlenkamp and Pavel Dolgirev for useful discussions.
This work was supported by the Swiss National Science Foundation (SNSF) under Grant Number 200021-204076. R.S. acknowledges support  from the Deutsche Forschungsgemeinschaft (DFG, German Research Foundation) under Germany's Excellence Strategy EXC 2181/1 - 390900948 (the Heidelberg STRUCTURES Excellence Cluster).

\section*{Appendix: Collective excitations and Chevy Ansatz}

\begin{figure*}[t]
    \centering
    \includegraphics[width=0.32\textwidth]{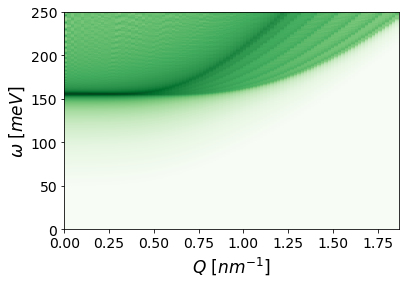}
    \includegraphics[width=0.32\textwidth]{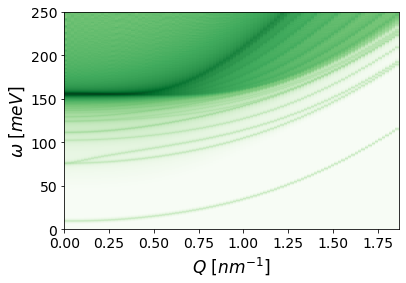}
    \includegraphics[width=0.32\textwidth]{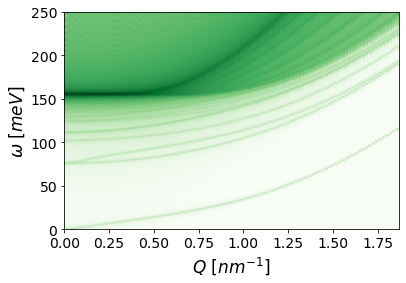}
    \caption{Density of states of the collective modes of momentum $Q$ for $d= L_1$ and $n \simeq 10^{12}$ cm$^{-2}$ in the absence of an impurity.
 In (a) the collective modes are computed from $H_\text{BCS}$, i.e.\ for non-interacting quasiparticles, and only the 2-quasiparticle continuum is correctly reproduced. In (b) instead
 one diagonalizes $\mathcal{A}(Q)_{kp}$, which expresses the action of $H_\text{el}$ in the tangent space to $|\text{EXI}\rangle$, and one can see the modes that sometimes are called Higgs and Bardasis-Schrieffer.
 The more refined calculation of panel (c) is instead based on the linearized equations of motion in the Gaussian manifold and captures the gapless Goldstone branch.
    The color scale is logarithmic in the three panels.}
    \label{fig:collective_modes}
\end{figure*}

For the Fermi polaron, the rationale behind writing the Chevy Ansatz as a superposition of the impurity times particle-hole excitations is that the neutral excitations  of a non-interacting Fermi sea are indeed particle-hole excitations.
% A good single-excitation polaron Ansatz should involve the neutral excitations of the system.
As we emphasized earlier, this is not the case for a BCS state: consequently, we will devote this Section to discuss the elementary excitations of the EXI
and we will prove that the BCS-Chevy Ansatz of Eq.~(\ref{eq:Chevy_ansatz})
describes correctly the scattering of the polaron with {\em most} of the neutral excitations of the bath.

The collective excitations on top of an EXI have been computed in \cite{wu2015theory, xue2020higgs}
in the language of time-dependent-Hartree-Fock theory.
Here we give a more geometric and modern presentation in terms of linearized dynamics within a manifold of Gaussian states, in the spirit of \cite{guaita2019gaussian, hackl2020}.
In this approach, the collective excitations can be obtained as variations of the Gaussian manifold
$ |\psi \rangle =  U_{\psi}| 0 \rangle$
:
\begin{equation}
    |\psi \rangle = 
    U_\text{BCS} \exp \left\{
    \sum_{kQ} \psi_k(Q) 
    b^\dagger_{k+Q} a_k
    - \psi^*_k(Q) 
    a^\dagger_{k} b_{k+Q}
    \right\} | 0 \rangle,
    \label{eq:Gaussian_el}
\end{equation}
where
we introduced the bare ``vacuum'' 
$a^\dagger a |0\rangle = |0\rangle, b |0\rangle = 0$
and the BCS variational ground-state
is given by 
$|\text{EXI}\rangle = U_\text{BCS} |0\rangle$
with 
\begin{equation}
    U_\text{BCS} =
    \exp \left\{ \sum_k \theta_k(b^\dagger_k a_k - 
    a^\dagger_k b_k ) \right\}
    % \times 
    % \exp \left\{ i \pi \sum_k b^\dagger_k b_k  \right\}
    .
\end{equation}
The quasiparticles are the rotated fermionic operators:
$(\alpha, \beta) =  U_\text{BCS} (a,b) U_\text{BCS}^\dagger$.

The tangent space to the manifold of
Eq.~(\ref{eq:Gaussian_el})
at the $|\psi\rangle$ point 
is spanned by 
\begin{equation}
   |kQ\rangle_\psi \equiv U_{\psi} b^\dagger_{k+Q} a_k
      | 0 \rangle.
\end{equation}
Since Gaussian states correspond to a K\"ahler manifold, the tangent space can be treated as vector space on the complex field~\cite{hackl2020}.
We note that equivalently, $|kQ\rangle_{\text{EXI}} =
\beta^\dagger_{k+Q} \alpha_k
   | \text{EXI} \rangle $
.

The dynamics within the Gaussian manifold is obtained by applying the Hamiltonian to the instantaneous state and projecting the result in the tangent space, in order to constrain the dynamics to the manifold, i.e.\
% \begin{equation}
%     i P_\psi \partial_t |\psi \rangle = 
%     P_|\psi \rangle H \psi 
% \end{equation}
% where $P_\psi$ is the projection operator.
\begin{equation}
    i  \partial_t \psi_k(Q) = 
    ~_\psi\langle kQ
    | H_\text{el} |\psi \rangle.
\end{equation}
These are in general highly nonlinear equations in the $\psi$'s and we stress that also the projector depends on $\psi$. Linearization of these equations yields two contributions: a normal term stemming from the variation of $|\psi\rangle$ 
and an anomalous term coming from the linearization of the projector $~_\psi\langle kQ
    |$.
    The matrix elements associated with the first term are nothing but the linearized Hamiltonian, i.e.\ the Hamiltonian in the tangent space.

An equivalent treatment is to write down  the Lagrangian 
associated to the manifold (\ref{eq:Gaussian_el}) 
and keep terms up to
 $O(\psi^2)$, to get
\begin{multline}
   \mathcal{L}_\text{el} =
   \sum_{kQ} \psi_k^*(Q) i \partial_t \psi_k(Q) - \sum_{kpQ} \mathcal{A}_{kp}(Q) \psi_k^*(Q) \psi_p(Q)
    -
    \\
    -
    \frac{1}{2}
    \mathcal{B}_{kp}(Q)
    \left[ 
    \psi^*_k(Q) \psi^*_p(-Q)
    +
     \psi_k(Q) \psi_p(-Q)
    \right] 
\end{multline}
where 
\begin{equation}
    \mathcal{A}_{kp}(Q) = \langle 
    \alpha_k^\dagger \beta_{k+Q} H_\text{el}
    \beta^\dagger_{p+Q} \alpha_{p}
    \rangle - 
     \langle H_\text{el}
    \rangle
\end{equation}
is the linearized Hamiltonian and 
\begin{equation}
    \mathcal{B}_{kp}(Q) = \langle 
     H_\text{el}
    \beta^\dagger_{k+Q} \alpha_{k} \beta^\dagger_{p-Q} \alpha_{p}
    \rangle
\end{equation}
gives the anomalous term. It is clear at this point that this is an equivalent approach to \cite{xue2020higgs}, where they report the same Lagrangian (the explicit expression for $\mathcal{B}_{kp}(Q)$ can be looked up there).

The Euler-Lagrange equations have the bosonic Bogoliubov form
\begin{equation}
    i\partial_t
    \begin{pmatrix}
    \psi_k(Q) \\
    \psi_{-k}^*(-Q)
    \end{pmatrix}
    =
    \begin{pmatrix}
        \mathcal{A}_{kp}(Q) & \mathcal{B}_{-kp}(-Q)\\
        -\mathcal{B}_{-kp}(-Q) & -\mathcal{A}_{kp}(Q)
    \end{pmatrix}
     \begin{pmatrix}
    \psi_p(Q) \\
    \psi_{-p}^*(-Q)
    \end{pmatrix},
    \label{eq:Euler-Lagrange}
\end{equation}
where this matrix is unitarily equivalent to 
\begin{equation}
    K%^\mu_{~\nu}
    =
    \begin{pmatrix}
        0 & \mathcal{A}_{kp}(Q) - \mathcal{B}_{-kp}(-Q)\\
        -\mathcal{A}_{kp}(Q) -\mathcal{B}_{-kp}(-Q) & 0
    \end{pmatrix}
\end{equation}
which is what is used in  \cite{guaita2019gaussian}.

The collective mode frequencies in this approximation are given by the eigenvalues of $K%^\mu_{~\nu}
$. However, the corresponding eigenvalues are {\em not} to be interpreted as the wavefunctions of the collective modes in the tangent space; this can already be understood from the fact that the dimension  of an eigenvector of $K%^\mu_{~\nu}
$ is twice the one of the tangent space.  
In other words, the collective modes do not live in the tangent space to the EXI state.
In our opinion, this is particularly clear in the geometric presentation used here.

In Fig.~\ref{fig:collective_modes}
we study the neutral excitations of the electronic system using this method, by plotting the resolvent or density of states (DOS) of the Bogoliubov matrix as a function of the momentum $Q$ of the collective excitation, using three different levels of approximation. Notice that the DOS does not tell anything about the sensitivity of a mode to a given probe.
In particular, in panel (a)  we just diagonalize $H_\text{BCS}$ in the tangent space and neglect the interactions between the quasiparticle.
In panel (b) we diagonalize the full 
$H_\text{el}$ in the tangent space, i.e.\ 
$\mathcal{A}_{kp}(Q)$. Scattering with these states is precisely what  is included in the Chevy Ansatz in the impurity problem.
Finally, in Fig.~\ref{fig:collective_modes}.c we diagonalize the full matrix of Eq.~(\ref{eq:Euler-Lagrange}).
In this way we recover a gapless Goldstone branch with acoustic dispersion at small momenta, a double branch that in the BEC limit corresponds to the 2p IX transition and in the BCS language is called Bardasis-Schrieffer mode~\cite{sun2020bash}, the 2s or Higgs branch~\cite{xue2020higgs}, a few other discrete modes and finally the two-quasiparticle continuum.
Interesting, the linearized Hamiltonian $\mathcal{A}_{kp}(Q)$ of panel (b) reproduces most of the DOS features; the Goldstone branch at small momenta however, is not captured and this approach yields a parabolic dispersion with a small gap that increases with density.
Diagonalizing $H_\text{BCS}$ instead yields only the continuum states.

For smaller densities, the gap in the DOS of $\mathcal{A}_{kp}(Q)$ gets very small (not shown); at the same time, in this limit the acoustic branch can be considered, for all practical purposes, to be parabolic. Consequently, the Chevy Ansatz Eq.~(\ref{eq:Chevy_ansatz}) is exact in the low density limit, provided that one keeps the interactions between the quasiparticles.

For higher densities, instead, it remains an open problem how much the gapless nature of the Goldstone branch would affect the polaron spectrum.

\bibliography{bibliography}

\end{document}